\documentclass[%
 reprint,
 amsmath,amssymb,
 aps,
]{revtex4-2}

\usepackage{graphicx}
\usepackage{dcolumn}
\usepackage{bm}
\usepackage{subfigure}

\begin{document}

\preprint{APS/123-QED}

\title{Successive one-sided Hodrick-Prescott filter with incremental filtering algorithm for nonlinear economic time series}

\author{Yuxia Liu}
\address{College of Engineering, Peking University, Beijing 100871, China.}

\author{Qi Zhang}
\thanks{zhangqi@uibe.edu.cn}
\address{School of Information Technology $\&$ Management, University of International Business $\&$ Economics, Beijing 100029, China.}

\author{Wei Xiao}
\address{College of Engineering, Peking University, Beijing 100871, China.
}%

\author{Tianguang Chu}%
\address{College of Engineering, Peking University, Beijing 100871, China.
}%




\date{\today}

\begin{abstract}
We propose a successive one-sided Hodrick-Prescott (SOHP) filter from multiple time scale decomposition perspective to derive trend estimate for a time series.
The idea is to apply the one-sided HP (OHP) filter recursively on the updated cyclical component to extract the trend residual on multiple time scales, thereby to improve the trend estimate.
To address the issue of optimization with a moving horizon as that of the SOHP filter, we present an incremental HP filtering algorithm, which greatly simplifies the involved inverse matrix operation and reduces the computational demand of the basic HP filtering. Actually, the new algorithm also applies effectively to other HP-type filters, especially for large-size or expanding data scenario. 
Numerical examples on real economic data show the better performance of the SOHP filter in comparison with other known HP-type filters.
\end{abstract}

\maketitle


\section{\label{sec:level1}Introduction}

Economic trends are intrinsic to determining the future development of the economy, and the proper filtering process enables us to extract the trend in economic time series and reduce noise \cite{alexandrov2012review}. 
A widely used method in econometrics for trend extraction is the Hodrick-Prescott (HP) filter, which splits a time series into the growth trend and the cyclical component, corresponding to low- and high-frequency components respectively \cite{toda2011income, jiang2019nonlinear}.
By its nature, the HP filter is a two-sided
filter, namely, it uses both future and past data to determine the current growth trend. 
This usually results in a smooth fitting of the trend.
To improve the trend estimate, \cite{Phillips2019} proposed the boosted HP (bHP) filter, which employs the HP filter to the residual cyclical component repeatedly to extract the leftover trend residual so as to alleviate over-smoothing effect.
Besides, the two-sided nature of the HP filter renders it unsuitable for forecasting and therefore \cite{stock1999forecasting} suggested a one-sided
HP (OHP) filter that only uses the currently available data in the HP filtering process. This effectively preserves the temporal ordering of the data and avoids over-smoothing of the current trend estimate with future data. To date the HP filter and its variants have found many applications in applied macroeconomic research \cite{Phillips2019, stock1999forecasting, yamada2018does}.

This paper intends to propose a successive one-sided HP (SOHP) filter from multiple time scale decomposition perspective. The motivation comes from the fact that an economic time series, such as stock series, is generically nonlinear and nonstationary, hence exhibiting multiple scale movement behavior \cite{kantz_schreiber_2003}. While the OHP filter can alleviate over-smoothing the trend to certain extend, there may be still leftover trend residual in the cyclical component separated in the filtering process. Hence, applying the OHP filter recursively on updated cyclical components allows us to extract the leftover trend residuals of higher frequencies successively and make modification to the trend estimate. 
This finally yields a multiple scale expansion of the trend. Numerical examples on the real data of the Standard \& Poor’s 500 (S\&P 500) index and the Shanghai Composite Index (SHCI) show that the SOHP filter evidently improves the trend estimates in comparison with the HP, bHP, and OHP filters. 

To facilitate the use of the SOHP filter, we also present an incremental HP filtering algorithm. The OHP filter is essentially an optimization problem with moving horizon \cite{stock1999forecasting}. Every time when a new observation is included, it has to repeat the entire procedure of the basic HP filter algorithm, which involves the inverse operation of a filtering matrix of the data size. Hence, in the process of the OHP filtering, one has to repeatedly compute the inverse of a matrix with ever increasing size. This is inconvenient and hampers the practical application of the SOHP filter. To circumvent the difficulty, we develop a novel incremental HP filtering algorithm that can generate the required inverse matrix for arbitrary finite size data in a recursive manner, with the only need of computing a third order initial inverse matrix once for all. This greatly reduces the computational demand of the basic HP filtering and facilitates application of existing HP-type filters to large-size or expanding data scinario.

\section{Incremental HP filtering algorithm}
\label{The bHP filter}

Recall that the HP filter \cite{Hodrick1997,  kovcenda2015elements} decomposes $l$ observations of a variable $y_t$ into the following form

\begin{equation}
	y_t = g_t + c_t  , \quad t = 1,2,\ldots,l, \label{app1}
\end{equation}
where $g_t$ represents the growth trend and $c_t$ the cyclical volatility, corresponding to the low- and high-frequency components of $y_t$, respectively.
The trend $g_t$ are determined by the minimization problem
\begin{equation}
	\min_{g_t} \left\{\sum_{t=1}^{l} (y_t - g_t)^2 + \eta \sum_{t=3}^{l} (g_t - 2g_{t-1} + g_{t-2})^2\right\} ,\label{app2}
\end{equation}
where $\eta\geq 0$ is a tuning parameter. Let $\boldsymbol{y}_l=[y_1, y_2, \cdots, y_l]^\top$, $\boldsymbol{g}_l=[g_1, g_2, \cdots, g_l]^\top$, and the tridiagonal matrix
\begin{equation*}
	F_l=\left[
	\begin{matrix}
		&1      & -2     & 1      & \cdots & 0 & 0  & 0   \\
		&\cdots &\cdots   &\cdots  &\cdots &\cdots   &\cdots  &\cdots \\
		&	0      & 0      & 0      & \cdots & 1 &-2  & 1    \\ 
	\end{matrix}
	\right]_{{(l-2)}\times{l}} ,
\end{equation*}
one can rewrite the problem (\ref{app2}) in a compact form as
\begin{align}
	&\min_{\boldsymbol{g}_l}\left\{||\boldsymbol{y}_l - \boldsymbol{g}_l||^2 + \eta ||F_l \boldsymbol{g}_l||^2 \right\}, \label{app3}
\end{align}
where $||\cdot||$ represents the $\ell_2$-norm of a vector.
The solution to the optimization problem (\ref{app3}) is given by
\begin{equation}
	\begin{split}
		&\boldsymbol{g}_l=S_l^{-1}\boldsymbol{y}_l, \label{app4}
	\end{split}
\end{equation}
and hence the cyclical part is
\begin{equation}
	\begin{split}
		&\boldsymbol{c}_l=(I_l-S_l^{-1})\boldsymbol{y}_l, \label{app90}
	\end{split}
\end{equation}
where $S_l=I_l +\eta F_l^\top F_l$ and $I_l$ is the $l\times l$ identity matrix.

Notice that the formulaes (\ref{app4}) and (\ref{app90}) explicitly depend on the length $l$ of the time series. 
When the length increases (as the case of the OHP filter), one has to compute the inverse of a new matrix $S_l$ in higher dimension. This would impede direct use of the formulaes in expanding sample or streaming data scenario. Actually, even for a time series of fixed length it is usually not desirable to take inverse operation for a matrix of large size directly. To address this issue, we propose an incremental algorithm for the HP filter as follows.
For $4 \le t \le l$, define
\begin{equation*}
	F_{t}=\left[
	\begin{matrix}
		\tilde{F}_{t-1} \\ 
		\boldsymbol{p}_{t}^\top 
	\end{matrix}
	\right]_{{(t-2)}\times{t}} ,
\end{equation*}
where $\tilde{F}_{t-1}=\left[F_{t-1} \quad \boldsymbol{0}\right]_{{(t-3)}\times{t}}$ and $\boldsymbol{p}_{t} = [0,\ldots,0,1,-2,1]^\top\in \mathbb{R}^{t}$, we have
\begin{align*}
	\label{app13}
	S_{t}=&I_{t} +\eta F_{t}^\top F_{t} \nonumber\\
	=&\left[
	\begin{matrix}
		S_{t-1} \quad& \\ 
		&\quad 1
	\end{matrix}
	\right] + \eta \boldsymbol{p}_{t} \boldsymbol{p}_{t}^\top.
\end{align*}
By the Woodbury matrix identity \cite{fornasier2011low}, we get the following recursive formula
\begin{equation}
	\label{app6}
	\left\{
	\begin{aligned}
		S_{t}^{-1}
		=&\left[
		\begin{matrix}
			S_{t-1}^{-1} \ &  \\
			&\  1
		\end{matrix}
		\right] - \left(\frac{1}{\eta} + \boldsymbol{p}_{t}^\top\boldsymbol{q}_{t}\right)^{-1}\boldsymbol{q}_{t}\boldsymbol{q}_{t}^\top,  \\
		S_{3}^{-1}
		=&\frac{1}{6\eta+1}\left[
		\begin{matrix}
			5\eta+1 \ &2\eta   \ &-\eta  \\
			2\eta   \ &2\eta+1 \ &2\eta  \\
			-\eta   \ &2\eta   \ &5\eta+1  \\
		\end{matrix}
		\right] ,
	\end{aligned}
	\right.
\end{equation}
where 
\begin{align*}
	\boldsymbol{q}_{t}=\left[
	\begin{matrix}
		S_{t-1}^{-1} \ & \\ 
		&\  1
	\end{matrix}
	\right]\boldsymbol{p}_{t},\quad \mathrm{for} \ 4 \le t \le l.
\end{align*}

Let $\boldsymbol{y}_{t}=[y_1, y_2, \cdots, y_{t}]^\top$, from Eqs. (\ref{app4}) and (\ref{app6}), the growth trend and the cyclical component of $\boldsymbol{y}_t$ are given by 
\begin{align}
	\boldsymbol{g}_{t}=&S_{t}^{-1}\boldsymbol{y}_{t} \nonumber\\
	=&\left[
	\begin{matrix}
		\boldsymbol{g}_{t-1}   \\
		y_{t}
	\end{matrix}
	\right]-\delta_t(g_{t-2}-2g_{t-1}+y_{t})\boldsymbol{q}_{t}, \label{app7}
\end{align}
and
\begin{align}
	\boldsymbol{c}_{t}=&\boldsymbol{y}_{t} - \boldsymbol{g}_{t}\nonumber\\
	=&\left[
	\begin{matrix}
		\boldsymbol{c}_{t-1}   \\
		0
	\end{matrix}
	\right]+\delta_t(g_{t-2}-2g_{t-1}+y_{t})\boldsymbol{q}_{t}, \label{app8}
\end{align}
respectively, where
\begin{equation*}
	\delta_t=\frac{1}{\frac{1}{\eta}+\boldsymbol{p}_{t}^\top \boldsymbol{q}_{t}}.
\end{equation*} 

\begin{figure}[h]
	\centering
	\includegraphics[scale=0.37]{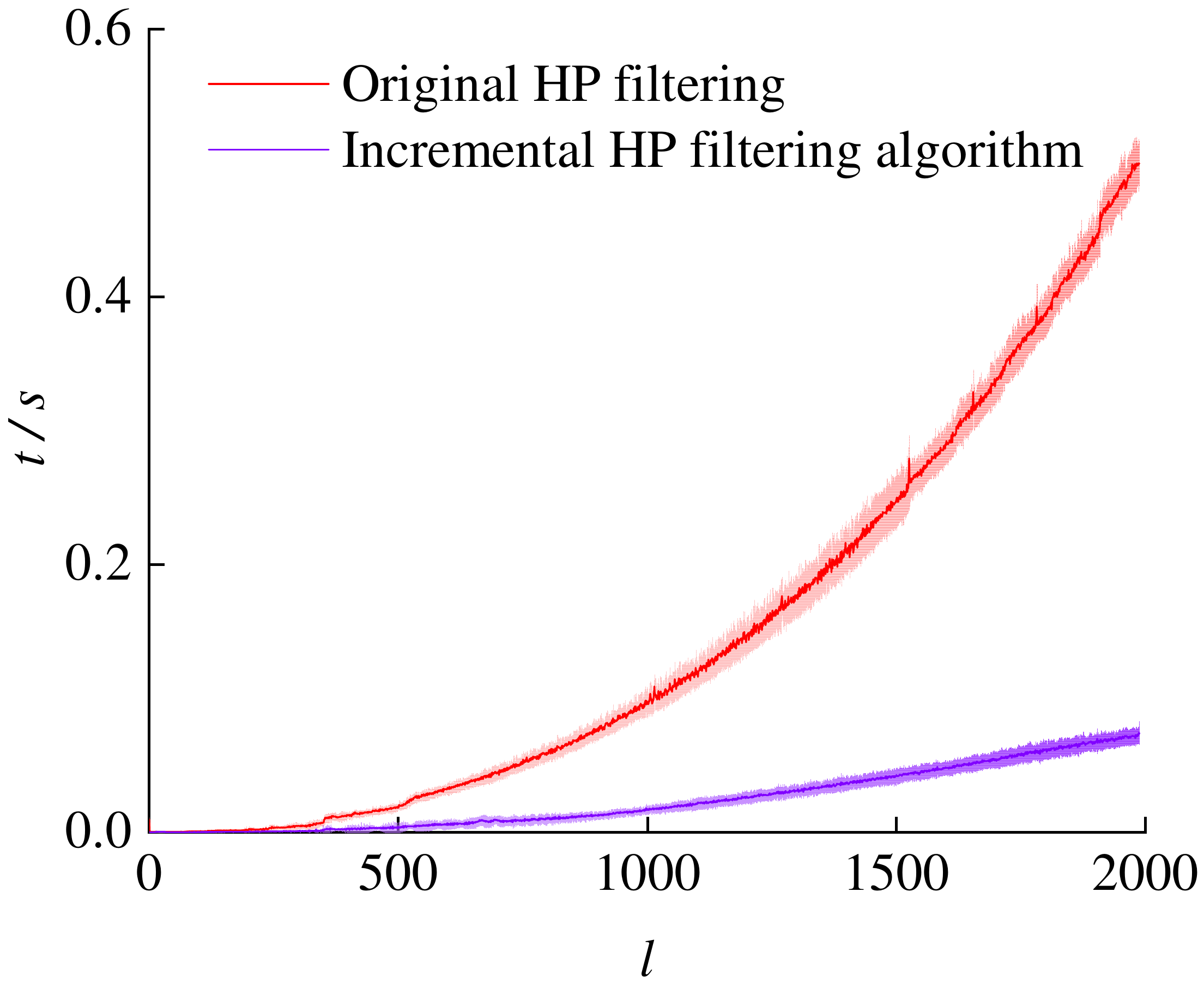}
	\caption{\label{fig:4} Time consumed by the original HP filtering and the incremental HP filtering algorithm.}
\end{figure}

We refer to Eqs. (\ref{app6})--(\ref{app8}) as an \emph{incremental HP filtering algorithm}, which merely requires evaluating a $3 \times 3$ initial inverse matrix and applies effectively to series data of fixed as well as expanding length in a recursive manner. It can be estimated that for a given length $l$, direct operation of the original HP filtering by Eq. (\ref{app4}) has the computational complexity of $\mathcal{O}(l^3)$, whereas the recursive formula (\ref{app7}) only has that of $\mathcal{O}(l^2)$. 
Fig. \ref{fig:4} displays a numerical test of the time consumption for the two algorithms at an Intel(R) Xeon(R) Gold 5118 CPU @2.30GHz 2.29GHz computer, with each algorithm repeated $100$ times and taking the average consumption time.
It clearly shows the time efficiency of our algorithm.

\section{SOHP filter}
\label{SOHP filter}

To motivate our study, we begin with the bHP filter proposed in \cite{Phillips2019}.
By definition, the bHP filter is a procedure of repeated applications of the HP filter on the residual $\boldsymbol{c}_{l}$ yielded in the last step. 
Thus after $n$ iterations of the filtering, the cyclical and trend components of the time series $\boldsymbol{y}_{l}$ are given by 
\begin{equation}
	\label{app9}
	\left\{
	\begin{aligned}
		\boldsymbol{c}_{l}^{(n)}=&(I_{l}-S_{l}^{-1})^n\boldsymbol{y}_{l}
		,  \\
		\boldsymbol{g}_{l}^{(n)}=&\left(I_{l}-(I_{l}-S_{l}^{-1})^n\right)\boldsymbol{y}_{l}.
	\end{aligned}
	\right.
\end{equation}
The resultant $\boldsymbol{g}_{l}^{(n)}$ includes trend residuals extracted from $\boldsymbol{c}_{l}^{(1)}, \cdots, \boldsymbol{c}_{l}^{(n-1)}$ and thus improves the trend estimate. Nevertheless, because it uses both future and past data in processing as the HP filter does, the bHP filter remains acting more like a smoother over the sample except for the end point. In addition, as noticed by \cite{stock1999forecasting}, introducing future data in estimating the trend is not reasonable for making macroprudential policy decisions. Hence, \cite{stock1999forecasting} proposed an OHP filter that applies the HP filter to the presently available sample only to avoid using future data at a given time.
For the time series $\boldsymbol{y}_{l}$, the growth trend given by the OHP filter can be expressed as

\begin{equation}
	\label{app11}
	\boldsymbol{g}^{O}_{l}=\sum_{t=1}^l \boldsymbol{a}_{t} \boldsymbol{g}_{t} \boldsymbol{e}(t),
\end{equation}
where $\boldsymbol{a}_{t}^{\top}=[0, \cdots, 0, 1]^{\top} \in \mathbb{R}^{t}$, $\boldsymbol{g}_{t}=S_t^{-1} \boldsymbol{y}_{t} \in \mathbb{R}^{t}$, and $\boldsymbol{e}(t)=[0, \cdots, 1, \cdots,0]^{\top} \in \mathbb{R}^{l}$ with the $t$-th value of 1.

Here we view the basic HP filter in Eqs. (\ref{app1}) and (\ref{app2}) as a slow- and fast-time scale decomposition technique and propose a new approach that employs the OHP filter in a recursive manner to  extract the volatility tendencies of the time series on different time scales. 
To be specific, we apply the OHP filter recursively to the residual cyclical component obtained in the last step of operation, following the procedure as below:

\textit{Step 1}. Apply the OHP filter to the data and obtain the trend $\boldsymbol{g}^{O{(1)}}_l=\boldsymbol{g}^{O}_l$ and the residual  $\boldsymbol{c}^{O{(1)}}_l=\boldsymbol{y}_l-\boldsymbol{g}^{O{(1)}}_l$;

\textit{Step 2}. Assume that for $i>1$, the cycle residual $\boldsymbol{c}^{O{(i)}}$ is available, apply the OHP filter to  $\boldsymbol{c}^{O{(i)}}_l$ to obtain higher order components $\boldsymbol{g}^{O{(i+1)}}_l$ and $\boldsymbol{c}^{O{(i+1)}}_l$, and repeat this operation on the updated cycle residual until certain stopping criterion (to be stated later) is met;

\textit{Step 3}. Sum up all the trend components  $\boldsymbol{g}^{O{(1)}}_l,\ldots,\boldsymbol{g}^{O{(n)}}_l$ of different scales successively obtained in $n$ iterations in \textit{Step 2} to get the finite expansion form of the trend estimate: 
\begin{equation}
	\begin{split}
		\overline{\boldsymbol{g}}_l=\sum_{i=1}^n \boldsymbol{g}^{O(i)}_l, \label{app10}
	\end{split}
\end{equation}
where $\boldsymbol{g}^{O(i)}_l$ denotes the $i$-th trend residul obtained in \textit{Step 2}. 

We will refer to the above approach as the \emph{successive one-sided HP (SOHP) filter}.

Analogous to the case of the bHP filter \cite{Phillips2019}, here we adopt the following stop criterion (SI) as the iteration stop condition for the SOPH in \textit{Step 2} : Let

\begin{equation}
	SI(n)=\frac{||\boldsymbol{c}^{O(n)}||_1}{||\boldsymbol{c}^{O(1)}||_1} +\frac{1}{l-2} \sum_{t=3}^l \frac{ \rm{tr}(\mathit{M_t^{(n)}})}{\rm{tr}(\mathit{I_t-S_t^{-1}})}, \label{app60}
\end{equation}
$||\cdot||_1$ represents $\ell_1$-norm of a vector, $\boldsymbol{c}^{O(n)}$ denotes the $n$-th cyclic residual yielded thereof, $\rm{tr}(\cdot)$ represents the trace of a matrix, and $M_t^{(n)} = I_t-(I_t-S_t^{-1})^n$; the proper number of iterations $n$ corresponds to the smallest $SI$ value.

Essentially, the SOPH filter is a procedure of extracting
volatility trends on different time scales to improve the trend estimation of a time series. 
Different from the HP and the bHP filiers, the use of the OHP filter in each round of iteration enables the SOHP filter avoiding to involve any future data in processing at any time. 
As a result, the SOHP filter can well retain the growth tendencies of a time series in comparison with the HP and the bHP filters, as to be seen later in our numerical experiments.

In the sequel we will incorporate the proposed incremental HP filtering algorithm into the bHP, the OHP, and the SOHP filters.
\section{Application to economic data}
\label{Applications to economic data}
We compare the performance of the HP, bHP, OHP, and SOHP filters using the real data of the S\&P 500 and the SHCI from Yahoo Finance \cite{Yahoo}.
The time series we analyzed consist of 885 observations including monthly data from January 1947 to September 2020 for the S\&P 500, and 279 observations including monthly data from July 1997 to September 2020 for the SHCI.
Let $y_t$ be the logarithm of these stock data,
we apply the four filters respectively to obtain trend estimates,  with the tuning parameter $\eta = 14400$ in Eq. (\ref{app2}) as suggested by \cite{FAVERO2001369, us2005optimal}.

According to our numerical experiments, the $SI$ in Eq. (\ref{app60}) assumes its minimum value at $n = 4$ and $n = 3$ for the S\&P500 and SHCI respectively.
Table \ref{bs1} shows the mean and the variance of the final $c_t$, as well as the number of iterations and the $SI$ value for the SOHP filter.  
\begin{table}[htbp]
	\caption{The SOHP filter with Bayesian-type information criterion.}
	\vspace{20pt}
	\centering
	\begin{tabular}{p{1.2cm}p{1.5cm}p{1.4cm}p{0.7cm}p{1.2cm}}
		\hline
		Index    & Mean  & Variance  &$n$  &$SI$ \\
		\hline   
		S\&P500  & 2.40e-4   &2.70e-3  &4 &0.8684 \\
		SHCI     & $-$5.01e-5  &1.75e-2 &3  &0.9659 \\     
		\hline       
	\end{tabular}
	\label{bs1}
\end{table} 

Fig. \ref{fig:5} plots the raw time series of the S\&P 500 and the SHCI, along with the corresponding trends estimated by different filters.  
Generally speaking, the performance of these filtering methods improves one by one from the HP filter to the SOHP filter. For instance, it can be seen from Fig. \ref{fig:5} (a) of the S\&P 500, where the shaded regions show the recessions dated by the NBER \cite{doi:10.1086/690241}, that the HP filter generates an over-smooth trend curve, erasing most of the sharp changes at the peaks and the troughs. This may attribute to the introduction of future data in the filtering process, which imposes on the filtered data some spurious patterns that are not part of the data generation process and cannot be identified from the real data, such as most of the unrecognized recessions in the shaded regions.
The bHP filter modifies the result of HP filter to certain extent by extracting the leftover trend residual from the updated cyclical component with the HP filter recursively.
Obviously, because of the two-sided nature of the primary HP filter, the bHP filtering process still intends to result in a smooth modification to the trend in general, incapable of recovering small scale tendencies in the trend. This is also clear by observing that most unrecognized recessions in the shaded remain unrectified by the bHP filter. Therefore, the improvement of the bHP filter is limited.
In contrast, the OHP filter successfully recognizes most of the recessions as well as some moderate scale tendencies in the trend.
The best result comes from the SOHP filter, which extracts more detailed growth tendencies on multiple time scales by recursively applying the OHP filter on the updated residual cycle term, thereby to recover evident sharp changes in the trend and all the recessions correctly. 
The case of the SHCI as shown in Fig. \ref{fig:5} (b) is likewise.

\section{Summary}
\label{Conclusion}
We have proposed the SOHP filter to improve the trend estimate of an economic time series by recursively extracting its growth tendencies of different scales. Numerical exercises on the S\&P 500 and the SHCI data demonstrate that in general the “two-sided approach” such as the HP and bHP filters tends to smooth the trend estimate and lose details in certain time scales. The “one-sided approach” like the OHP and SOHP filters is capable of retaining main inherent growth tendencies of the time series. By comparison, the SOHP filter outperforms the other three due to its recursive manner that allows for extracting growth trends of different scales. Meanwhile, the proposed incremental HP filtering algorithm greatly simplifies the involved invers matrix operation and  reduces the computational demand. A salient feature of the new algorithm is that it is suitable for fixed-length data as well as expanding or streaming data, making the HP-type filters ready for modern data-rich environments in economic research.

\begin{figure*}[htbp]
	\centering
	\subfigure[The S\&P 500 ]{
		\includegraphics[scale=0.6]{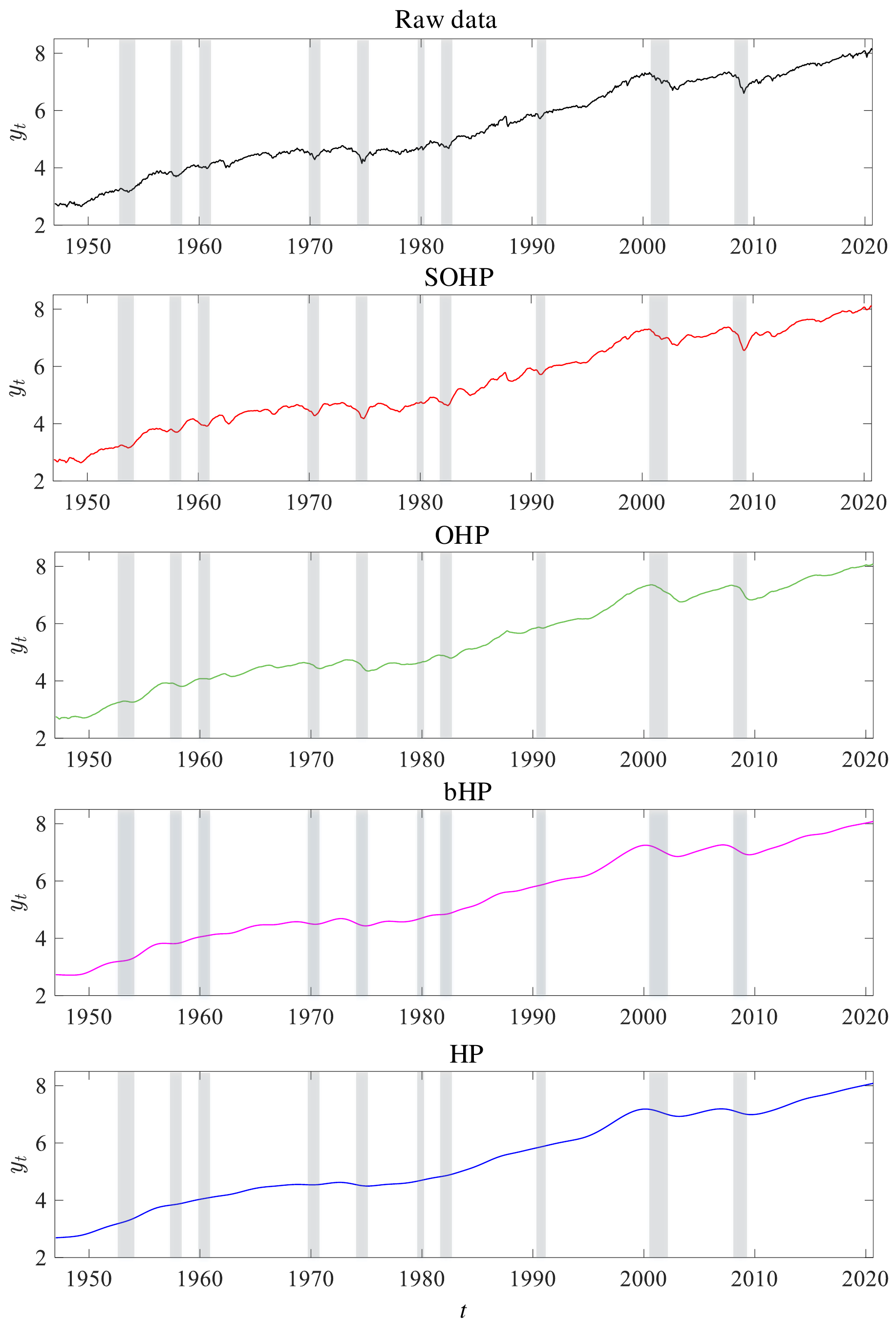}
	}
\end{figure*}
\addtocounter{figure}{-1}       
\begin{figure*} 
	\addtocounter{figure}{1}      
	\centering 
	\subfigure[The SHCI ]{
		\includegraphics[scale=0.6]{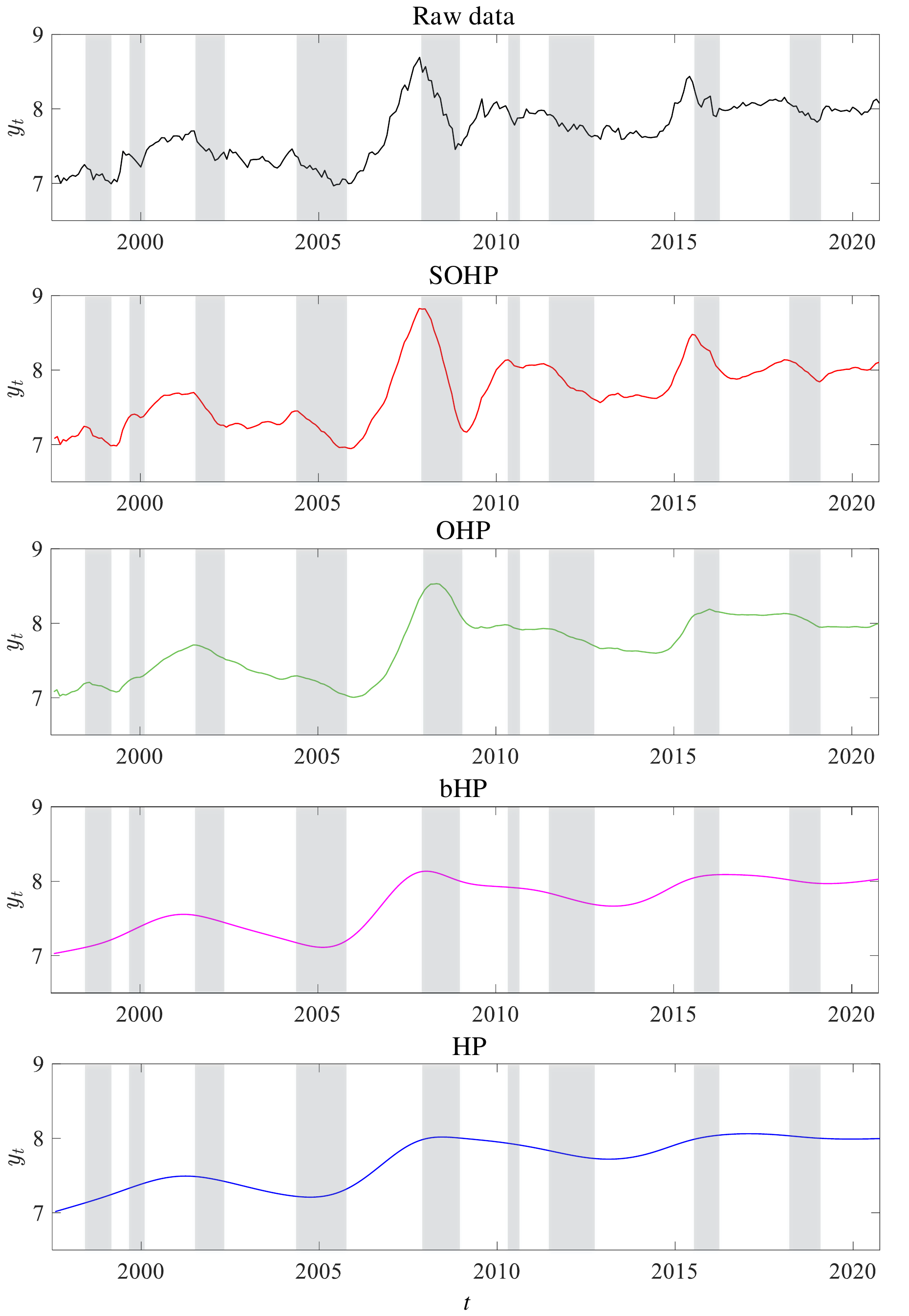}
	}	
	\caption{\label{fig:5} Trend components with respect to different filters.}
\end{figure*}

\bibliography{apssamp}

\end{document}